# Performance of Vehicle-to-Vehicle Communication using IEEE 802.11p in Vehicular Ad-hoc Network Environment

Vaishali D. Khairnar[1] and    Dr. Ketan Kotecha[2]

[1]Research Scholar Institute of Technology Nirma University Ahmadabad

`khairnar.vaishali3@gmail.com`

[2] Director Institute of Technology Nirma University Ahmadabad

`director.it@nimauni.ac.in`

## ABSTRACT

*Traffic safety applications using vehicle-to-vehicle (V2V) communication is an emerging and promising area within the ITS environment. Many of these applications require real-time communication with high reliability. To meet a real-time deadline, timely and predictable access to the channel is paramount. The medium access method used in 802.11p, CSMA with collision avoidance, does not guarantee channel access before a finite deadline. The well-known property of CSMA is undesirable for critical communications scenarios. The simulation results reveal that a specific vehicle is forced to drop over 80% of its packets because no channel access was possible before the next message was generated. To overcome this problem, we propose to use STDMA for real-time data traffic between vehicles. The real-time properties of STDMA are investigated by means of the highway road simulation scenario, with promising results.*

## KEYWORDS

CSMA, STDMA, V2V, VANET etc.

## 1. INTRODUCTION

The new emerging applications for enhancing traffic safety found within the vehicular ad-hoc network environments which can be classified as real-time system. Existing vehicle-to-vehicle safety systems together with new cooperative systems using wireless data communication between vehicles which can potentially decrease the number of accidents on the highway road in India i.e. transmit the messages within deadlines. In addition, requirements on high reliability and low delay are imposed on wireless communication system [1]. For example Lane departure warning messages merge assistance and emergency vehicle routing are all examples of applications [2]. Information that is delivered correctly, but after the deadline in a real-time communication system, is not only useless, but can also have severe consequences for the traffic safety system. This problem is pointed out in [3-4]. In most cases, the extremely low delays required by traffic safety applications, the need for ad-hoc network architectures support direct vehicle-to-vehicle communication.  The IEEE 802.11p standard intended for vehicle-to-vehicle ad-hoc communication in high speed vehicular network environments [5], which states amongst other things that multiple data/packet exchanges should be completed within 50 milliseconds of time frame.

The original IEEE 802.11, intended for WLAN, has two drawbacks within its MAC technique CSMA; it can cause unbounded delays before channel access as well as collisions on the channel. The MAC protocol decides who has right data/packet to transmit next on the shared communication channel. In CSMA, the node first listens to the channel an if the channel is free for certain amount of time period, then the node transmits data/packets directly with the implication that another node can have conducted the exact same procedure, resulting in a collision on the channel. So a node can experience very long channel delays due to the risk of the channel being busy. CSMA is used by the whole IEEE 802.11 family as well as its wired counterpart IEEE 802.3 Ethernet. One of the reasons for the success of both WLAN and Ethernet is the straightforward implementation of the standard resulting in reasonable priced equipment. Due to this WLANs and the Ethernet are often applied to other domains than they originally are designed for. Even though CSMA is unsuitable for real-time vehicle-to-vehicle communication because of the unbounded channel access delays, Ethernet has played its way communication scene where many real-time systems are found. The problems with MAC protocol can be solved here by introducing more network equipment, such as switches and routers, and thereby reducing the number of nodes competing for the shared channels, i.e. breaking up collision domains. But in wireless domain, there is no such easy solution since the wireless channel has to be shared by all users. When the CSMA algorithm is applied in the wireless domain, an interferer could easily jam a geographical area, the nodes in this area would defer their access even though there is no real-time data traffic present. A wireless CSMA system is thus more susceptible to interference since no access will occur as long as activity is detected on the channel.

The IEEE 802.11p, also known as Dedicated Short-Range Communication (DSRC), intended for vehicular ad-hoc networks (VANETs). Currently this is the only standard with support for direct vehicle-to-vehicle (V2V) communication [6]. The original DSRC standards, which are found in Europe, Japan and Korea, are more application-specific standards containing the whole protocol stack with a physical (PHY), a MAC and an application layer. They are intended for hot spot communication such as electronic toll collection systems. The PHY in 802.11p and its capabilities have been treated in several articles [7-9]. The PHY mainly affects the reliability (error probability) of the system; however, if we do not get channel access the benefits of the PHY cannot be exploited. VANET will use CSMA as its MAC method, despite its inability to support real-time deadlines. The argument is that the problems with CSMA are most pronounced at high network loads, and traffic smoothing can be introduced to keep the data traffic at an acceptable level. Traffic smoothing is typically used in centrally controlled networks in restricted geographical areas [7-10]. A vehicular ad-hoc network (VANET) is neither a restricted geographical area, nor can it be made predictable by a central controller due to its highly dynamic characteristics and requirements on low delay. Traffic smoothing only reduces the average delay and the main problem with unbounded worst case delay remains. The problem with potentially unbounded channel access delays when using CSMA could be to use STDMA (self-organizing time division multiple access), a decentralized, yet predictable, MAC protocol with a finite channel access delay, making it suitable for real-time ad-hoc vehicular networks. STDMA algorithm is already in commercial use in the system called automatic identification system, where it focuses on collision avoidance between vehicles.

This paper analyzes the particular vehicle-to-vehicle communication real-time requirements on the MAC protocols when used in vehicular ad-hoc network environment system for low-delay traffic safety applications. The two different MAC protocols are evaluated by means of computer simulations: the MAC protocol in 802.11p, CSMA and a solution potentially better suited for decentralized real-time system, namely STDM, First an introduction to the concept of MAC protocol is given with the functionality descriptions of CSMA and STDMA. Next, the system model is detailed and results from the simulator are displayed. The paper is finally concluded with a discussion and conclusions regarding the two examined MAC protocols in the context of the traffic safety applications on highway road.

## 2. MEDIUM ACCESS CONTROL

Vehicular ad-hoc network (VANETs) is a spontaneous, unstructured network based on direct vehicle-to-vehicle (V2V) communication and its topology is changing constantly due to high mobility of vehicle nodes on highway road. In VANET it is harder to deploy a MAC scheme that is relying on a centralized controller, for example TDMA (time division multiple access), FDMA (frequency division multiple access) or CDMA (code division multiple access). In a centralized infrastructure-based network, a base station/access point is responsible for sharing the resources among the users, thereby enabling guaranteed QoS for time-sensitive data traffic. The idea of having a vehicle node that could act as a central control unit in a distributed VANET is not appealing because of the high mobility vehicle nodes. The central unit would not remain central for long and constantly changing the central unit would require much data exchange and negotiation among the vehicle nodes. The negotiation can be expected to incur excessive delay and once a decision is made it is likely to already be outdated. MAC scheme that does not require a central control unit is CSMA, where each vehicle node starts by listening to the wireless channel and transmits only if the channel is ideal. This scheme is easily deployed in a distributed network, but has one big disadvantage; the vehicle nodes could experience unbounded delays due to constantly sensing a busy channel during high utilization periods. This is not acceptable in real-time environments.

Real-time systems such as traffic safety applications, call for a deterministic MAC protocol. We define a deterministic MAC protocol method to be a scheme for which the time from channel access request to channel access has a finite upper bound. In the following we will evaluate CSMA and STDMA in this respect.

### 2.1 IEEE 802.11p/DSRC Protocol

The IEEE 802.11p standard (WAVE) emerges from the allocation of the Dedicated Short Range Communications (DSRC) spectrum band in the United States and the work done to define the technology to be used in this band. There are two types of channels in DSRC, all of them with a 10 MHz width: the control channel (CCH) and the service channel (SCH). The CCH is restricted to safety communications only, and the SCHs are available both for safety and non-safety use. Applications for vehicular communications can be placed in three main categories - traffic safety, traffic efficiency and value-added services (e.g. infotainment/business) [11]. In 1999, the U.S. Federal Communication Commission (FCC) allocated these 75 MHz of spectrum at 5,850-5,925

GHz to be used exclusively for vehicle-to-vehicle and infrastructure-to-vehicle communications. The main objective is to enable public safety applications in vehicular environments to prevent accidents (traffic safety) and improve traffic flow (traffic efficiency).

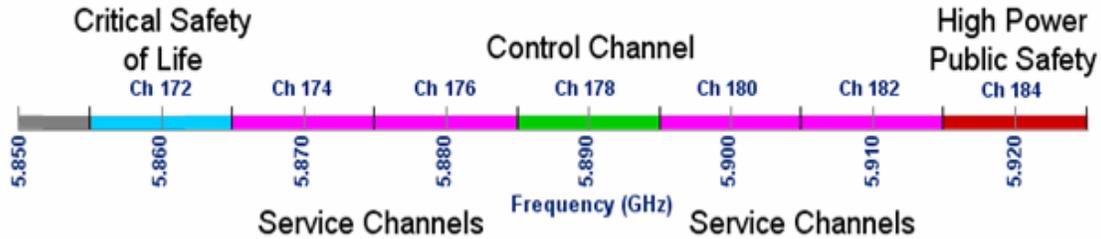

Fig. 1 DSRC spectrum band and channels.

In Europe, the spectrum allocated by the ETSI for cooperative safety communications has a range 5,875 - 5,925 GHz. It is divided into traffic safety (30 MHz) and traffic efficiency (20 MHz). In the traffic safety spectrum, two SCHs and one CCH are allocated. In the traffic efficiency two SCHs are allocated [11]. As stated before, WAVE has its origins in the standardization of DSRC as a radio technology.

WAVE is fully intended to serve as an international standard, which is meant to: describe the functions and services required by WAVE stations to operate in VANETs, and define the WAVE signaling technique and interface functions that are controlled by the IEEE 802.11 MAC. WAVE is an amendment to the Wireless Fidelity (WiFi) standard IEEE 802.11 [18]. It is inside the scope of IEEE 802.11a, which is strictly a PHY and MAC level standard. In other words, IEEE 802.11p is an adaptation of the IEEE 802.11a protocol to vehicular situations, such as: rapidly changing environment. With a short time frame transactions required, and without having to join a Basic Service Set (BSS) (Peer-to-Peer (P2P) and ad-hoc networks). WAVE is only a part of a group of standards related to all layers of protocols for DSRC-based operations as can be seen in Fig 2. In this paper, we are only going to focus on IEEE 802.11p.

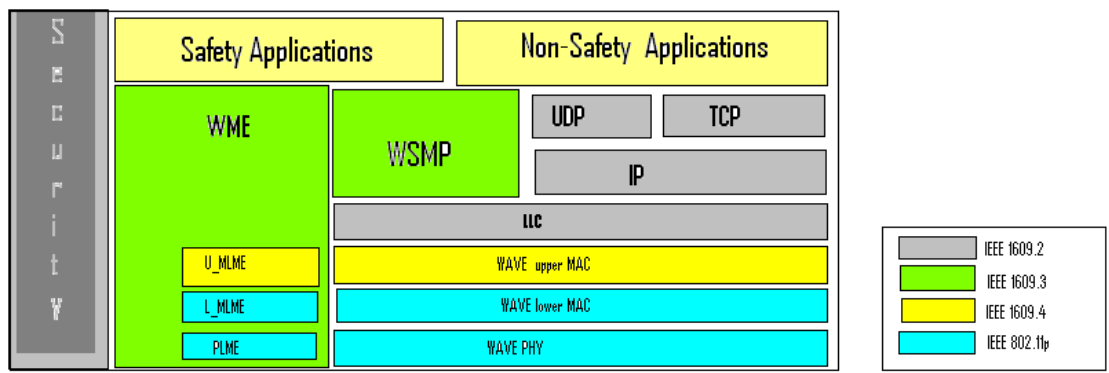

Fig. 2 DSRC standards and communication stack.

ETSI has defined two central types of messages: Central Access Messages (CAMs), and Decentralized Environmental Notification Messages (DENM). IEEE 802.11p has adopted these two types of messages. As a reminder, CAMs are broadcast packets sent periodically at a concrete heartbeat rate. A CAM packet contains information about the stating vehicle speed, position and driving direction of the transmitter [12]. DENMs are event-driven and application specific messages, which are sent on emergency cases. They are triggered in case of a hazard and are continuously broadcasted until this hazard disappears [12]. Every node in the network manages the information received by remote nodes, as well as the data generated by the own vehicle. All data is contained in a database called Local Dynamic Map (LDM). CAMs and DENMS are used to update the LDM of each vehicle with the information gathered from the rest of nodes of the network.

### 2.1.1 Data frame format

A procedure carried out by the PLCP sub-layer is the convergence procedure, in which it converts the actual data frame being sent, named PLCP Service Data Unit (PSDU) into the PLCP Protocol Data Unit (PPDU). In this procedure, the preamble and header are appended to the PSDU to obtain the PPDU. The preamble consists of 12 training symbols, 10 of which are short and are used for establishing automatic gain control, diversity selection and the coarse frequency offset estimate of the carrier signal. The receiver uses 2 long training symbols for channel and fine frequency offset estimation. It takes up to 16 ms to train the receiver after first detecting a signal on the RF medium. The header, also called the SIGNAL field of the PPDU frame, is always transmitted at 6 Mbps using BPSK modulation. It contains information about the transmission data rate and type of modulation (BPSK, QPSK, 16QAM or 64 QAM) in the RATE field and the length in number of octets of the PSDU that the MAC is currently requesting to transmit in the LENGTH field; as well as a parity bit (Parity field), based on the first 17 bits, and a Tail field with all bits set to 0. The PSDU itself is pre-pended with the Service field, with the first 7 bits as zeros to synchronize the descrambler in the receiver and the remaining 9 bits reserved for future use and set to all 0s, and appended with the Tail field and Pad Bits field, which are the number of bits that make the DATA field a multiple of the number of coded bits in an OFDM symbol (48, 96, 192, or 288). The Service field, PSDU, Tail field and Pad Bits field form the DATA field of PPDU frame. The IEEE 802.11 PPDU frame format is shown in Fig 3. [13]

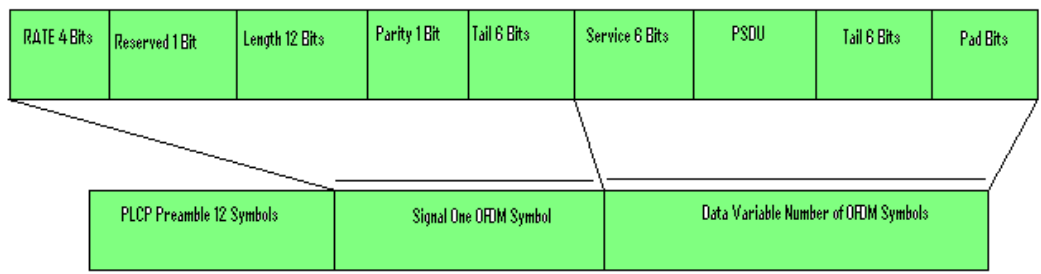

Fig 3: The IEEE 802.11 PPDU frame format

### 2.1.2 OFDM (Orthogonal Frequency Division Multiplexing)

The transmission of data is based on Orthogonal Frequency Division Multiplexing (OFDM) technique. OFDM divides the available band into K sub-bands or sub-carriers, which are separated a frequency bandwidth F. From this perspective, OFDM is similar to Frequency Division Multiple Access (FDMA). However, in FDMA all subcarriers require spectral guard intervals in order to prevent interferences between closely allocated subcarriers. OFDM uses the spectrum much more efficiently than FDMA since it makes all the subcarriers orthogonal to each other. This way, it is possible to have the subcarriers all together as close as possible and prevent any interference amongst them. Orthogonality of the sub-carriers means that an integer multiple of cycles is contained in each symbol interval in every different subcarrier. Thus the spectrum of each subcarrier has a null at the central frequency of each of the other subcarriers. This attenuates the problems of overhead carrier spacing and guard interval allocation required in FDMA. In IEEE 802.11p there are 64 subcarriers, but only the 52 inner subcarriers are used. 48 out of these 52 actually contain the data and 4 of them, called pilot subcarriers, transmit a fixed pattern used to mitigate frequency and phase offsets at the receiver side. Each of these 48 data subcarriers can be modulated, as explained before, with BPSK, QPSK, 16QAM or 64QAM. In combination with different coding rates, this leads to a nominal data rate from 6 to 54 Mbps if full clocked mode with 20MHz bandwidth is used [14]. However, a change has been done in terms of sampling rate, for the adaptation of IEEE 802.11a to IEEE 802.11p: in IEEE 802.11p a channel of 10 MHz bandwidth is used. This way, the guard interval is long enough to prevent Inter-Symbol Interference (ISI) caused by multipath channel during the transmission and hence it fits the high-speed vehicular environment that characterizes the VANETs. The parameters in the time domain are doubled, compared to the parameters in IEEE 802.11a [11]. In Table 1 some of these parameters are shown.

| Parameters | IEEE 802.11a | IEEE 802.11p | Changes |
|---|---|---|---|
| Channel bandwidth | 20 MHz | 10 MHz | Half |
| Bit rate (Mbps) | 6,9,12,18,24,36,48,54 | 3,4.5,6,9,12,18,24,27 | Half |
| Modulation Mode | BPSK, QPSK, 16QAM, 64QAM | BPSK, QPSK, 16QAM, 64QAM | No change |
| Number of subcarriers | 52 | 52 | No change |
| Symbol duration | 4µs | 8µs | Double |
| Guard Interval Time | 0.8 µs | 1.6 µs | Double |

Table 1: Comparison of PHY parameters in IEEE 802.11a and IEEE 802.11p

## 2.1.3 The transmitter

The binary data that is to be sent over the wireless medium, which is the PSDU, is encoded and modulated. The resulting coded data string is constantly being assigned to a certain complex number in a signal constellation and groups of 48 of these complex numbers are mapped to OFDM subcarriers. The operation in the assembler block is, mainly, to insert 4 pilot subcarriers among the 48 data subcarriers and form the OFDM symbol. In the next block, the OFDM sub-carriers are converted to the time domain using the Inverse Fast Fourier Transformation (IFFT) preparation time domain signal with circular extension of itself to generate the cyclic prefix. In the last stage in the transmitter, all the OFDM symbols are appended one after the other to form the PSDU and appended again with the PLCP preamble, the PLCP header (SERVICE field of the PPDU) and the fields Service, Tail and Pad Bits. This way the PPDU is obtained and is ready for transmission [14]. The block diagram of the transmitter is depicted on**.**

## 2.1.4 The channel

All types of wireless communications, the medium is the radio channel between transmitter and receiver. The signal is propagated through different paths, that can be either Line-Of-Sight paths (LOS) or Non-Line-Of-Sight (NLOS) between transmitter and receiver. In each of these paths, the signal can suffer from reflections, scattering and diffractions by different objects during its itinerary. These are just few of the conditions that can affect the multipath communications in this medium, and have a big impact on the propagation of the VANETs. Most times it is very complicated to take into account all of the adversities found this medium, therefore simplified model channels are used. For VANETs, these models must take into account the existence of multiple propagation paths and the high relative velocities among nodes.

Two key parameters that are directly affected by the channel conditions are:

Signal-to-Noise Ratio (SNR) is broadly defined as the ratio of the desired signal power to the noise power. This ratio indicates the reliability of the link between the receiver and the transmitter.

$$SNR[dB] = PowerRcvd[dB] - 10 * log10(noise) \qquad \textbf{Eqn. 1}$$

Signal-to-Interference-to-Noise Ratio (SINR) is defined as the ratio of the desired power to the noise power plus the interferences generated by other transmitters close to the analyzed one, which are also considered as noise for the receiver.

$$SINR[dB] = PowerRcvd[dB] - 10 * log10( \ (PowerInt) + noise) \qquad \textbf{Eqn. 2}$$

## 2.1.5 The receiver

In the receiver part, for the adaptation of IEEE 802.11a to IEEE 802.11p, some required improved performances have been introduced in the receiver to avoid cross channel interferences from adjacent channels [11]. The first block in the receiver is the Serial-to-Parallel (S/P), in which the signal is divided in blocks of samples and the DATA field is separated from the Preamble and SIGNAL fields of the PPDU. Both DATA and Preamble are demodulated with the Fast Fourier Transform (FFT) algorithm. After that, the channel coefficients are estimated and based on them, the equalizer compensates the fading effects introduced by the channel and transmits the samples to the decoder. Finally the received and decoded binary data stream is compared to the transmitted one, in order to calculate the error ratio statistics [14]. The block diagram of the transmitter is depicted.

## 2.2 MAC Layer

For the adaptation of IEEE 802.11a to IEEE 802.11p, no changes in the MAC layer have been done. The MAC protocol used in 802.11p is the same as in 802.11a, the Enhanced Distributed Channel Access (EDCA), which is an enhanced version of the basic access mechanism in IEEE 802.11 using Quality of Service (QoS) [20-22].

## 2.2.1 Overview of MAC services

### 2.2.1.1 Data service

This service provides peer entities in the LLC (Local Link Control) MAC sub-layer with the ability of exchanging MSDUs (MAC Service Data Units) using the underlying PHY-layer services. This delivery of MSDUs is performed in an asynchronous way, on a connectionless basis. By default, MSDU transport is based on best-effort. However, the QoS facility uses a Traffic Identifier (TID) to specify differentiated services on a per-MSDU basis. There are no guarantees that the MSDUs will be received successfully. Broadcast and multicast transport is part of the asynchronous data service provided by the MAC layer. Due to the characteristics of the wireless medium, broadcast and multicast MSDUs may experience a lower QoS, compared to that of unicast MSDUs. In our simulations, only broadcast MSDUs are sent and received and no acknowledgement is used, and the vehicles also called stations, are nQSTAs (non-QoS STAtions), this means that no QoS is used since all the transmitted messages are the same type (CAMs). [15] The encapsulation of a MSDU inside a MPDU (MAC Protocol Data Unit), which becomes the PDSU when processed at a PHY layer level.

### 2.2.1.2 MSDU ordering

In nQSTAs, the ones simulated in this paper, there are two service classes within the data service. By selecting the desired service class, each LLC entity initiating the transfer of MSDUs is able to control whether MAC entities are or are not allowed to reorder those MSDUs at reception. In an nQSTA, the MAC does not intentionally reorder MSDUs. If a reordering happens, the sole effect of this (if any), for the set of MSDUs received at the MAC service interface of any single STA, is a change in the delivery order of broadcast and multicast MSDUs originating from a single source STA address. If a higher layer protocol using the data service cannot tolerate this possible reordering, the optional Strictly Ordered service class should be used. [15] No reordering of MSDUs takes place in our simulations [25-27].

## 2.3 MAC sub-layer functional description

### 2.3.1 MAC architecture

The MAC architecture can be described as shown in Fig 4 as providing the Point Coordination Function (PCF) and Hybrid Coordination Function (HCF) through the services of the Distributed Coordination Function (DCF). The HCF is composed by the HCF contention-based channel access also called Enhanced Distributed Channel Access (EDCA), the HCF Controlled Channel Access (HCCA) and the Point Coordination Function (PCF) [23, 24].

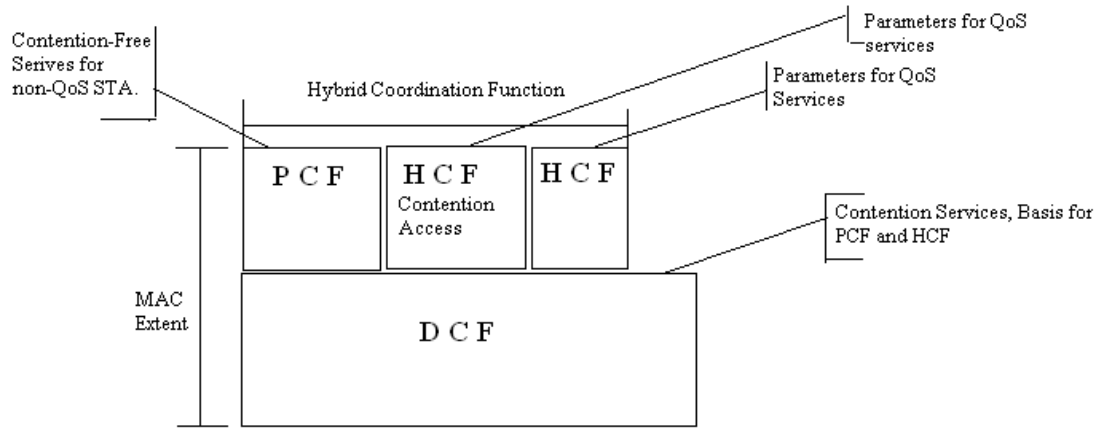

Fig :4 MAC architecture

### 2.3.1.1 D C F

DCF is the fundamental MAC technique in the IEEE 802.11 standard. It employs an access function performed by the CSMA/CA algorithm and a collision management function carried out by the binary exponential back-off procedure.

### 2.3.1.2 P C F

The original IEEE 802.11 standard defines another coordination function in the MAC layer. It is only available in structure mode networks, where the nodes are interconnected through at least one AP in the network. This mode is optional and only very few APs or Wi-Fi adapters actually implement it. The coordinator block is called Point Coordinator (PC). In the scope of this paper, PCF is not used because we are not simulating an infrastructure network (with Access Points (APs)), but a VANET, where all nodes are peers: not only the vehicles but also the road-side infrastructure behave as peers in a VANET.

### 2.3.1.3 H C F

HCF is a coordination function that enables the QoS facility. It is only usable in networks that make use of QoS, so it is only implemented in the QSTAs. The HCF combines functions from the DCF and PCF with some enhanced, QoS-specific mechanisms and frame subtypes to allow a uniform set of frame exchange sequences to be used for QoS data transfers. The HCF uses both a controlled channel access mechanism, HCCA, for contention-free transfer and a contention-based channel access method mechanism, EDCA.

### 2.3.1.4 H C C A

HCCA works similarly to PCF. It uses a QoS-aware centralized coordinator, called a Hybrid Coordinator (HC), and operates under rules that are different from the PC of the PCF. HCCA is generally considered the most advanced (and complex) coordination function. With the HCCA, QoS can be configured with great precision. QSTAs have the ability to request specific transmission parameters which allow advanced applications to work more effectively on a Wi-Fi network.

## 2.4 Operation of the CSMA/CA algorithm

Since we are not dealing with different type of messages, all the packets sent by the nodes have the same priority and the QoS enhancements explained before that EDCA adds are not needed. We give to these packets the highest priority and for that reason, we use AIFS = 58 µs and CW = CWmin = 3. Furthermore, we will not suffer in our simulations from virtual collisions, but only from real collisions [15]. In addition, all the messages sent are broadcasted and because of that we do not make use of the SIFS concept neither. We are dealing with nQSTAs, so HCF is not present in our simulations. What it is really of interest in this paper from the IEEE 802.11p MAC layer are the CSMA/CA algorithm and the exponential back off procedure found in DCF.

The CSMA/CA procedure according to IEEE 802.11p, it is, in the broadcast situation with periodic data traffic (CAM packets), is presented in Fig 5.

The transmitter node starts by listening to the channel activity during an AIFS amount of time (which in our simulations is 58 µs). If after this time, the channel is sensed free, the packet is transmitted. After that, the node checks if a new packet from the upper layers is ready to be transmitted, and when there is one, it performs the same action to transmit the new packet. If during AIFS, the channel is busy or becomes busy, then the node gets a random back off value, generated from an exponential distribution, by multiplying the integer from [0..CW] with the slot time 13 _s obtaining 0, 13, 26 or 39 _s. This value will be decreasing every time the node waits for an AIFS and senses the channel free. When the back off value gets to 0, then the packet can be transmitted. While the node is getting its back off value decreased, it keeps on checking constantly if a new packet was generated in the upper layers and is ready to be transmitted. When that happens, the old packet is dropped, and the node starts again with the whole transmission protocol.

## 2.5 Self-Organizing Time Division Multiple Access (STDMA) MAC Layer Algorithm

The STDMA algorithm, invented in [16, 17], is already used in commercial applications for surveillance, i.e., the Automatic Identification System (AIS) used by ships and the VHF data link (VDL) mode 4 system used by the avionics industry. Traditional surveillance applications for airplanes and ships are based on ground infrastructure with radar support. Radar has shortcomings such as the inability to see behind large obstacles or incorrect radar images due to bad weather conditions. By adding data communication based on STDMA, more reliable information can be obtained about other ships and airplanes in the vicinity and thereby accidents can be avoided.

Since STDMA is so successful in these systems, it is interesting to investigate if it can manage a more dynamic setting such as a vehicular network. STDMA is a decentralized MAC scheme where the network members themselves are responsible for sharing the communication channel. Nodes utilizing this algorithm, will broadcast periodic data messages containing information about their position. The algorithm relies on the nodes being equipped with GPS receivers. Time is divided into frames as in a TDMA system and all stations are striving for a common frame start. These frames are further divided into slots, which typically corresponds to one packet duration. The frame of AIS and VDL mode 4 is one minute long and is divided into 2250 slots of approximately 26 ms each. All network members start by determining a report rate, i.e., how many position messages that will be sent during one frame. Then follows four different phases; *initialization, network entry, first frame,* and *continuous operation*. During the *initialization,* a node will listen to the channel activity during one frame length to determine the slot assignments. In the *network entry* phase, the node determines its own transmission slots within each frame according to the following rules: (*i*) calculate a nominal increment (*NI*) by dividing the number of slots with the report rate, (*ii*) randomly select a nominal start slot (NSS) drawn from the current slot up to *NI*, (*iii*) determine a selection interval (SI) of slots as 20% of *NI* and put this around the NSS according to Fig. 1, (*iv*) now the first actual transmission slot is determined by picking a slot randomly within SI and this will be the nominal transmission slot (NTS). If the chosen NTS is occupied, then the closest free slot within SI is chosen. If all slots within the SI are occupied, the slot used by a node furthest away from oneself will be chosen. When the first NTS is reached in the super frame, the node will enter the third phase called the *first frame*. Here a nominal slot (NS) is decided for the next slot transmission within a frame and the procedure of determining the next NTS will start over again. This procedure will be repeated as many times as decided by the report rate (i.e., the number of slots each node uses within each frame), Fig. 6.

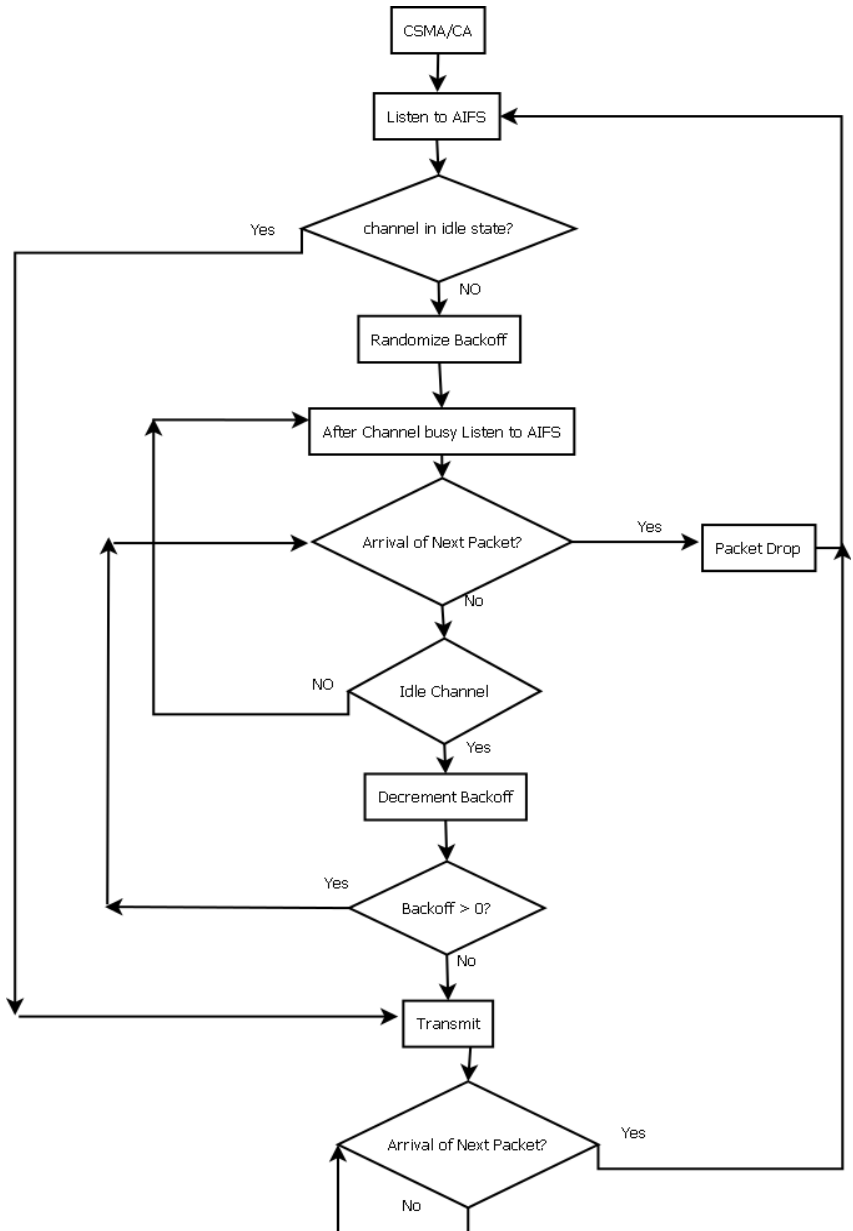

Fig 5: The CSMA / CA procedure according to 802.11p

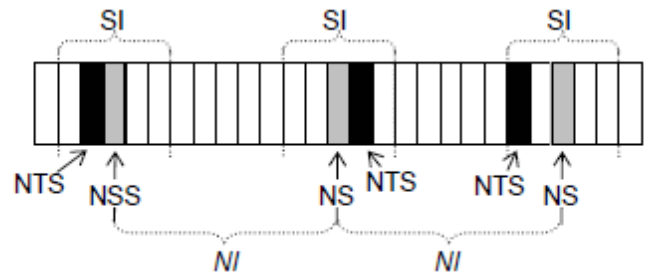

Fig 6. The STDMA algorithm in the first frame phase.

After the first frame phase (which lasts for one frame) when all NTS were decided, the station will enter the *continuous operation* phase, using the NTSs decided during the *first frame* phase for transmission. During the *first frame* phase, the node draws a random integer *n* {3,...,8} for each NTS. After the NTS has been used for *n* frames, a new NTS will be allocated in the same SI as the original NTS. This procedure of changing slot after a certain number of frames is to cater for network changes, e.g., two nodes using the same NTS which were not in radio range of each other when the NTS was chosen could have come closer and will then interfere. The STDMA relies on the position information sent by other network members and it will not work without this.

### 2.5.1 Continuous operation phase

The last phase is called continuous operation phase. Here, a new concept is introduced, the n reuse factor. Every message in a slot has an n value related to it, which decreases within every transmission. When n gets to 0, then the message has to be reallocated in a new slot within the same SI as the former slot. If all of the slots are busy, then the procedure is the same as in the second phase. Apart from a reallocation, a new n factor is assigned to the new NTS location. This factor is used to cater with changes in the network topology. When a node enters the same transmission range of another node, and both of them have a message allocated in the same slot within the frame, it will cause a co-located transmission and in case they are close to each other packets from both co-located transmitters might be lost by the receiving nodes. Without the use of the n reuse factor, they would be suffering a collision every time until they get out of the same range of transmission. The situation changes when one of them gets its n reuse factor value to 0, so its message has to be reallocated to a new slot avoiding from that moment, suffering a collision with the other node. The n reuse factor adds flexibility to STDMA, very important since we are dealing with VANETs, whose nodes are constantly moving. The continuous operation phase is depicted as a flow diagram in Fig 7 [18-19, 30].

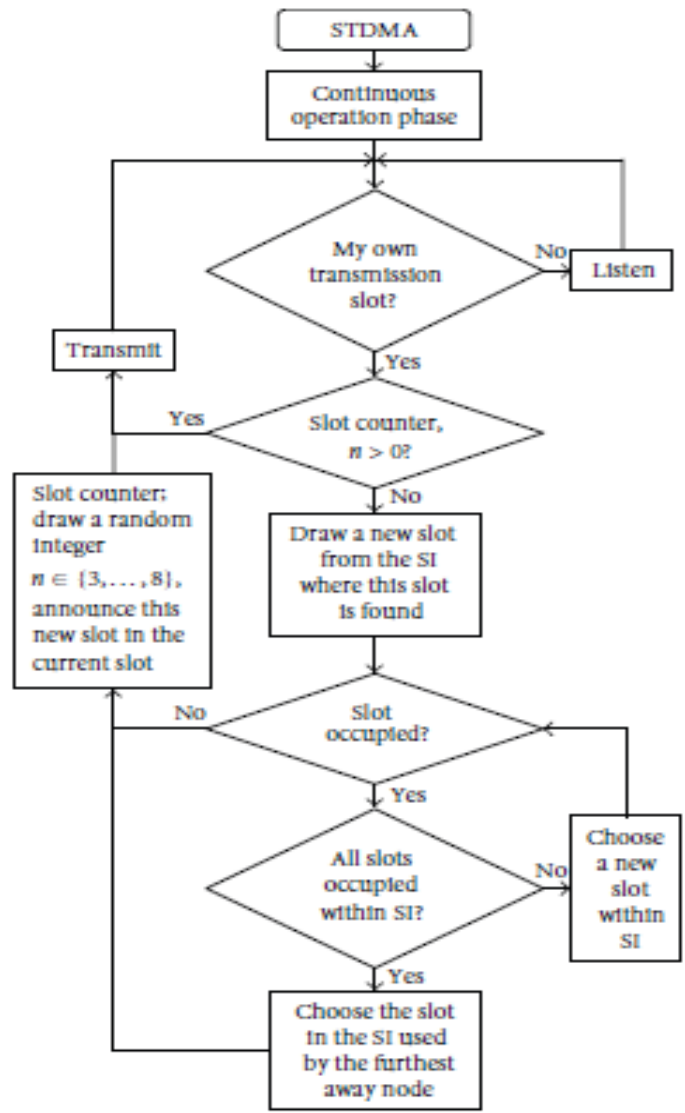

Fig 7: The continuous operation phase of STDMA

## 3. SIMULATIONS

The real-time properties of the system, the interesting issue here is how the two MAC protocols will influence the capability of each sending vehicle node to timely deliver data/messages packets, i.e. meeting real-time deadlines. We are dealing with an uncontrolled network since the number of network vehicle nodes cannot be determined in advance as we are considering vehicles are controlled by humans. On the highway road, the highest relative speeds are found and this causes the network topology to change often and more rapidly. If a traffic accident occurs, many vehicles could be gathered in a small geographic area implying troubles with access to the shared wireless communication channel for individual vehicle nodes. Here we are studying the MAC channel access delay for time-driven location messages; we are not considering the reception of messages at the vehicle nodes at this time.

The promising emerging application within VANET is a cooperative awareness system such as the automatic identification system for the ships, where the vehicles will exchange location messages with each other to build up a map of its surrounding and use this for different traffic safety efficiency application [27]. Routing is highly mobile networks and is also dependent on location rather than specific addresses when trying to find ways through the network. Therefore, time-driven location messages are likely to be of uttermost importance in future vehicular networks. Consequently, we have also chosen to use broadcasted, time-driven location messages as the data traffic model in the simulator. Many traffic safety systems will rely on vehicles periodically broadcasting messages containing their current state (e.g., current location, speed, average speed, distances travelled, total distance etc). We have developed a simulator using Open Street Map, eWorld, SUMO version 0.10.3 (traffic simulator), NS-2 version 2.34 (Network Simulator) and TraNs version 1.2 (Intermediate simulator between SUMO and NS2) also we require Gnu plot/Xgraph/Excel to plot the graphics presentation (Fig 11 for Simulation Flow diagram) where each vehicle sends a location message according to a predetermined range of 5 or 10 Hz. Simulations has been conducted both for the CSMA of 802.11p as well as for the proposed STDMA algorithm. The vehicle traffic scenario is a Mumbai-Pune Highway Road of 120 kilometer (km) i.e. 12000 meter with 3 lanes in each direction (i.e. total 6 lanes including both the directions) Fig 8.

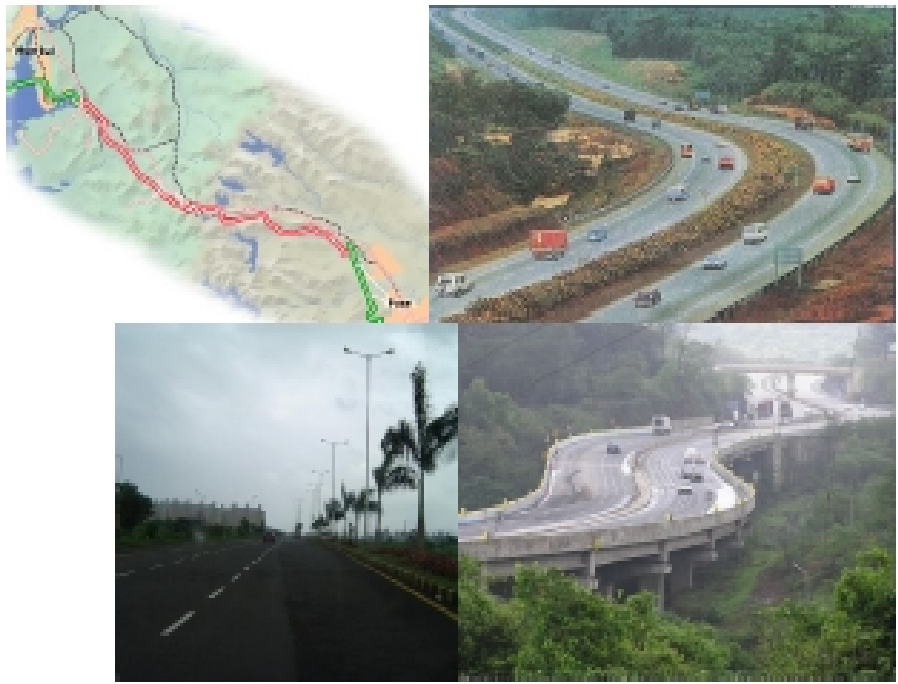

Fig 8. Scenario of Mumbai-Pune Highway Road.

The Mumbai-Pune Highway Road scenario is chosen because here the highest relative speeds (i.e. min 80 km/h to max 120 or above km/h) in vehicular environments are found and hence it should constitute the biggest challenge for the MAC layer. The vehicles are entering each lane of the highway road according to a Poisson process with a mean inter-arrival time of 3 seconds (consistent with the 3-second-rule used in Sweden, which recommends drivers to maintain a 3 second spacing between vehicles). The speed of each vehicle is modeled as a Gaussian random variable with different mean values for each lane; 83 km/h, 108 km/h and 130 km/h, and a standard deviation of 1 m/s. For simplicity we assume that no overtaking is possible and vehicles always remain in the same lane. There is no other data traffic in addition to the heartbeat broadcast messages. The channel model is a simple circular transmission model where all vehicles within a certain sensing range will sense and receive packets perfectly. The simulated sensing ranges are 500 m and 1000 m. We have tried to focus on how the two MAC methods perform in terms of time between channel access requests until actual channel access within each vehicle node. Three different packet lengths have been considered: 100, 300 and 500 byte. The shortest packet length is just long enough to distribute the location, direction and speed, but due to security overhead, the packets are likely longer [28]. The transfer rate is chosen to be the lowest rate supported by 802.11p, namely 3 Mbps. Since all vehicles in the simulation are broadcasting, no ACKs are used. Table 2. Contains a summary of the simulation parameter settings.

The channel model is a simple circular sensing range model, Fig. 10, in which every vehicle node within the sensing area receives the message perfectly. Note the vehicle nodes could be exposed to two concurrent transmissions, where transmitters TX1 and TX2 are sending at the same time since the transmitters cannot hear each other: the receivers RX1, RX2, and RX3 in Fig. 13 will then experience collisions of the two ongoing transmissions, unless some sort of power control or multiuser detection is used. However, since the focus of this simulation is to characterize the MAC channel access delay $T_{acc}$, problems such as exposed and hidden terminals are not addressed here. As soon as the vehicle nodes enter the highway road, they will start to transmit after an initial random delay of between 0 to 100 ms. The simulation has been carried out with three different packet lengths: N=100, 300 and 500 bytes and two different sensing ranges 500 and 1000 meters. The sensing range of 1000 meters was chosen because of 802.11p states that communication ranges up to 1000 meters must be supported and the different data/packet lengths are chosen because of the security issues. It is very important that range messages can be trusted since many traffic safety applications will be depending on these. One way to handle the security issue is to use a digital signature being approximately 125 bytes [29] and in worst case this signature must be included in every packet. Therefore, 500 byte data packets should be the worst case length of range packets including a signature of 125 bytes, together with the header, trailer and location data/message.

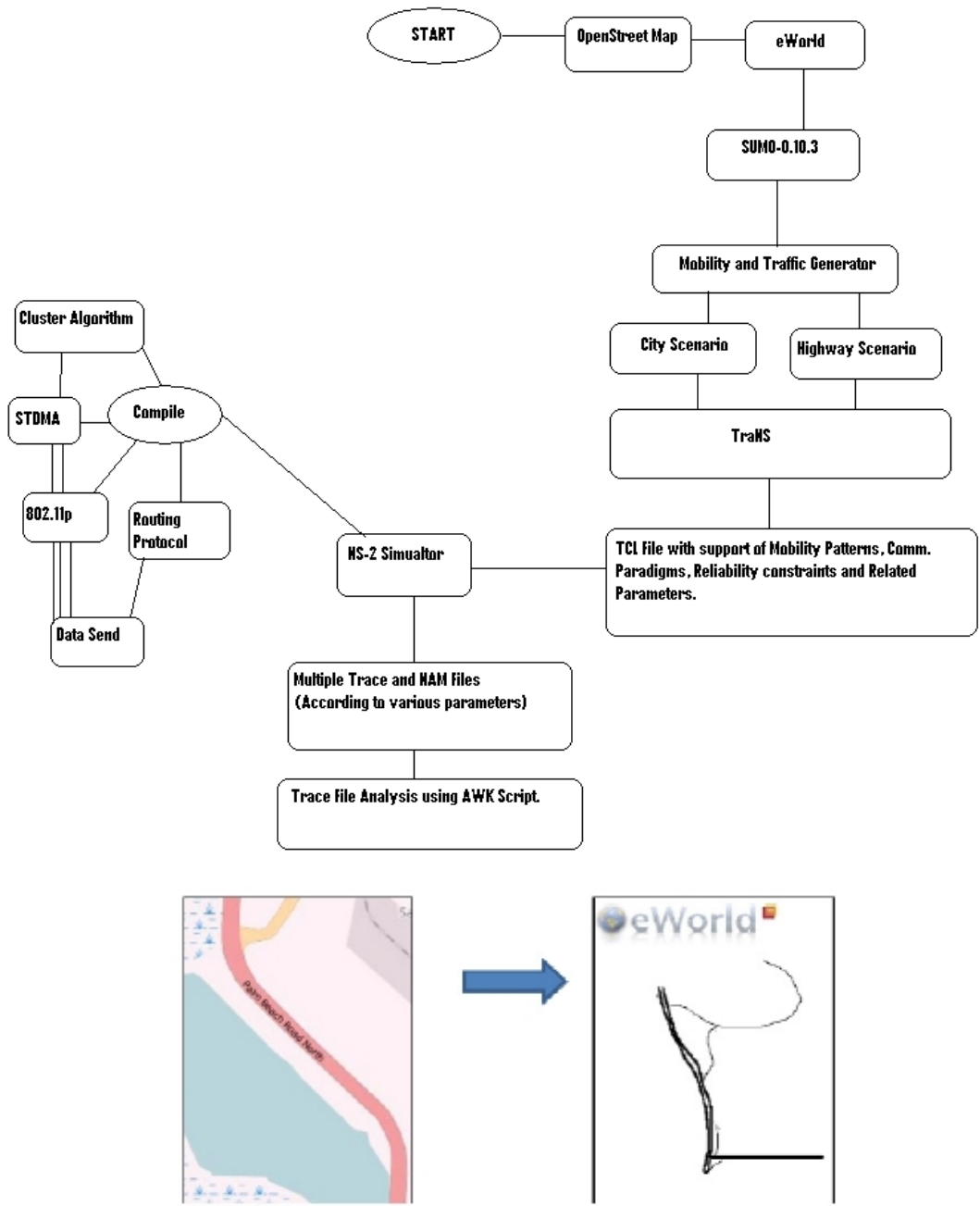

Fig 9. Simulation Flow Diagram

| Parameter | Value |
|---|---|
| Start-point of Highway Road | Panvel |
| End-Point of Highway Road | Pune |
| Simulation Time | 1 hour 30 mins. (In-time 6.30 am & Out-time 8.00 am) |
| Length of Highway Road | 120 Km or 12 000 m |
| Traffic direction | 2 ways |
| Number of Lanes in each direction | 6 lanes ( 3 in each direction ) |
| Vehicle type | Cars, Private vehicles, Buses, Trucks etc. |
| Number of Vehicle Nodes on Highway | 1200 |
| Speed of Vehicle nodes | 40 – 120 km/h, |
| Communication Protocol | 802.11p and STDMA |
| Traffic type | UDP |
| Packet sending frequency | 5 Hz, 10 Hz |
| Packet length | 100 bytes (Ratio 30% vehicles), 300 bytes (Ratio 40% vehicles) and 500 bytes (Ratio 30% vehicles). |
| Transfer Rate | 3 Mbps |
| Slot time, $T_{slot}$ | 9µs |
| SIFS, $T_{SIFS}$ | 16µs |
| CWmin | 3 |
| CWmax | Not used |
| Communication Range | 250 meter, 500 meter |
| Backoff Time, $T_{Backoff}$ | 0,9,18,27 µs |
| AIFS (listening time before sending) CSMA parameter | 34 µs (highest priority) |
| STDMA frame size | 1 s |
| No of slots in the STDMA frame | 3076 slots (100 byte packets), 1165 slots (300 byte), 718 slots (500 byte) |

Table 2. Simulation parameter setting for Mumbai-Pune Highway Road scenario simulation

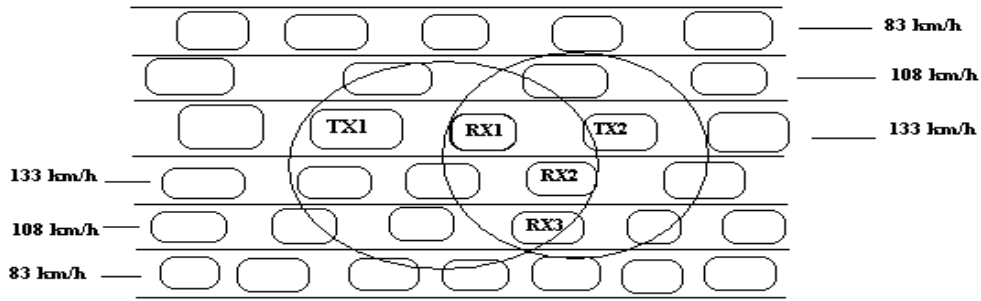

Fig. 10: Simulation Setup.

We used in our CSMA simulations, all vehicles use the MAC method of 802.11p, and hence each vehicle must listen before sending and backoff if the channel is busy or becomes busy during the automatic identification system. A broadcast packet will experience at most one backoff procedure due to the lack of ACKs in a broadcast system. The contention window will never be doubled since at most one failed channel access attempt can occur. In the Table 3, parameters used in the simulation of 802.11p are listed. Since all data traffic in our simulation scenario has the same priority, only the highest priority automatic identification system and $CW_{min}$ have been used and therefore all transmitters will have the same $T_{AIFS}$ value (34µs). The backoff time is the product of the slot time, $T_{slot}$, and a random integer uniformly distributed in the interval [0,3] implying four possible backoff times, $T_{backoff}$: 0,9,18 and 27 µs respectively. In Fig. 5, a flow diagram presents the CSMA procedure in the broadcast situation with periodic location messages from every vehicle node. The "Next data packet arrived?" box tests if the new location message has arrived from the layer above the MAC layer, in which case the old data packet awaiting channel access is outdated and will be dropped.

The STDMA algorithm found in automatic identification system cannot be used right away since the dynamics of a vehicular ad-hoc network and a shipping network are quite different. The automatic identification system is using lower frequencies for transmission to reach further away and the ships need to know much further ahead about ships in the vicinity to take the right decisions early on. There is a natural inertia inherent in a shipping system that is not present in the vehicular environment. i.e. braking a truck and turning a ship in an emergency situation are two very different tasks. We have much shorter time frames to work within the vehicular ad-hoc network environment. Both MAC protocols used in the simulation are assumed to use the same physical layer from 802.11p. The frame duration, $T_{frame}$, in our simulated STDMA scheme has been set to 1 second and the number of slots is changed inside the frame to cater for different packet lengths. A transfer rate, R of 3 Mbps has been used and this rate is available with the PHY layer of 802.11p, which has support for eight transfer rates in total where 3 Mbps is the lowest. This choice is made since the system under consideration requires high reliability rather than high throughput, and the lowest transfer rate has the most robust modulation and coding scheme. In the STDMA simulations, the vehicles will go through three phases: initialization, network entry and first frame, before it ends up in the continuous operations. The phases are described in Fig.10 the continuous operation phase. The vehicle stays in the continuous phase after it has been through the other three. STDMA always guarantees channel access even when all slots are occupied within an SI, in which case a slot belonging to the vehicle node located furthest away will be selected. The time parameters involved in the simulation are selected from the PHY specification of 802.11p. The CSMA transmission time, $T_{CSMA}$, consists of an AIFS period $T_{AIFS}$ in 34µs, a 20 µs preamble, $T_{preamble}$, and the actual data packet transmission, $T_{packet}$. The STDMA transmission time, $T_{STDMA}$, which is the same as the slot time, consist of two guard times, $T_{GT}$, of 3 µs each, $T_{preamble}$, $T_{packet}$, and two SIFS periods, $T_{SIFS}$ of 16 µs each derived from the PHY layer in use. SIFS stands for short interframe space and accounts for the transceiver to switch from sending to receiving state and vice versa plus the MAC processing delay. The total transmission time for CSMA is

$$T_{CSMA} = T_{AIFS} + T_{preamble} + T_{packet} \qquad \textbf{Eqn 3}$$

and the total transmission time for STDMA is

$$T_{STDMA} = 2\ T_{GT} + 2\ T_{SIFS} + T_{preamble} + T_{packet} \qquad \textbf{Eqn 4}$$

In Table 3 the different timing parameters are shown for different data packets.

| Packet Length N (byte) | $T_{packet}$ ($\mu$s) | $T_{CSMA}$ ($\mu$s) | $T_{STDMA}$ ($\mu$s) | No. of slots |
|---|---|---|---|---|
| 100 | 267 | 321 | 325 | 3076 |
| 300 | 800 | 854 | 858 | 1165 |
| 500 | 1333 | 1387 | 1391 | 718 |

Table 3: The transmission times for CSMA and STDMA.

We have assumed that all the vehicles nodes in the system are perfectly synchronized with each other in both MAC protocol scenarios and that in the STDMA case they are also aware of when the frame starts and how many time slots it contains. The delay that takes to a packet sent from the transmitting vehicle until it is decoded by the receiving vehicle at the MAC layer level. This measure shows not only the delay, but also the reliability of the messages since it takes into account the interference at the MAC level caused by other vehicles.

This delay is expressed as:

$$T_{MM} = Tca + Tp + Tdec \qquad \textbf{Eqn 5}$$

At the receiver side, to be a packet candidate to be decoded and sent to higher layers, it should have arrived within 100 ms, which is the maximum allowed delay at the receiver vehicle for CAM messages to be considered.

The values analyzed from this performance indicator are the mean values of TMM for a concrete message transmitted by a concrete vehicle to all of the receiver vehicles.

# 4 RESULTS

We evaluate CSMA and STDMA in terms of channel access delay. Simulations have been carried out with the parameter settings in Table 2, considering 10 different scenarios. Data from the simulations have been collected only when the Mumbai-Pune highway road was filled with vehicles i.e. during peak hours. The results from all 10 simulated scenarios using CSMA are shown in Table 4 where the numbers represent the data packet drops in percent. A data packet is dropped or discarded by the vehicle node when the next data packet is generated. The old data packet is dropped because a newer data packet with more accurate location data has arrived from the application within the vehicle node. We consider the channel access delay to be infinite for dropped data packets.

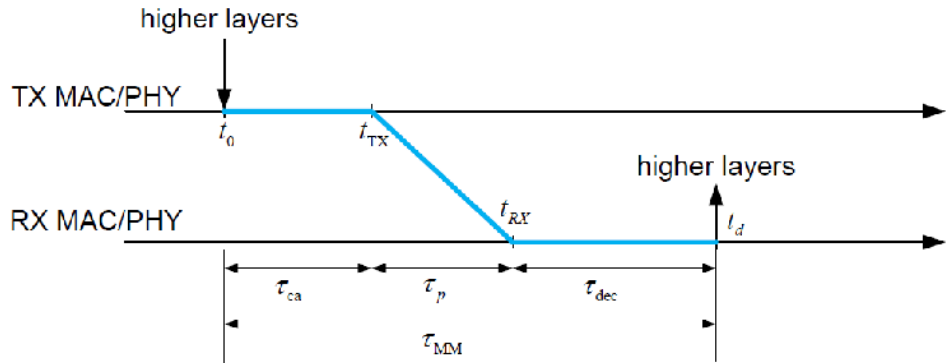

Fig.11 Mac-to-Mac (E2E) Delay

| CSMA | | Sensing range | | | |
|---|---|---|---|---|---|
| | | 500 meter | | 1000 meter | |
| Data Packet Rate | | 5Hz | 10 Hz | 5Hz | 10 Hz |
| Packet length | 100 byte | 0% | 0% | 0% | 0% |
| | 300 byte | 0% | 0% | 0% | 36% |
| | 500 byte | 0% | 23% | 34% | 54% |

Table 4: Packet drops on average for different data traffic scenarios.

From Table 4 it can be seen that, if 500 byte long data packets are sent every 100ms and the sensing range is 1000 meters, only 48% of the channel access request will result in actual channel access for 802.11p. But, this value is averaged over all transmissions made by all vehicles in the system which means that certain nodes experience an even worse situation. In Fig. 12, the best and worst performance experienced by a single user is depicted together with the average for all users in the system. In the worst case, a vehicle node achieves successful channel access only 16% of the time i.e. 80% of all generated packets in this vehicle node are dropped. When the sensing range is 1000 meters, a vehicle node will complete for the channel with approximately 230 other vehicle nodes.

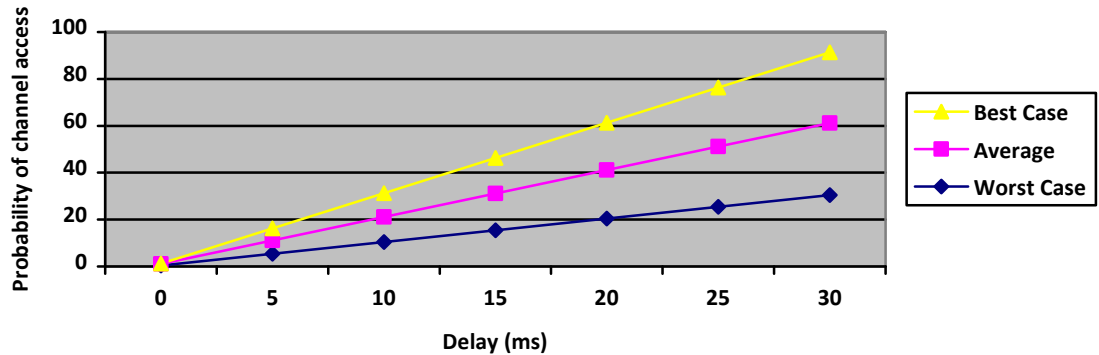

Fig. 12: Channel access delay in CSMA with a sensing range of 1000m, report rate 10HZ and packet length 500 byte.

In fig. 13, the results from a sensing range of 500 m are depicted, and the worst-case vehicle nodes are experiencing data packets drops 55%. In this scenario, approximately 115 vehicle nodes are competing for channel access.

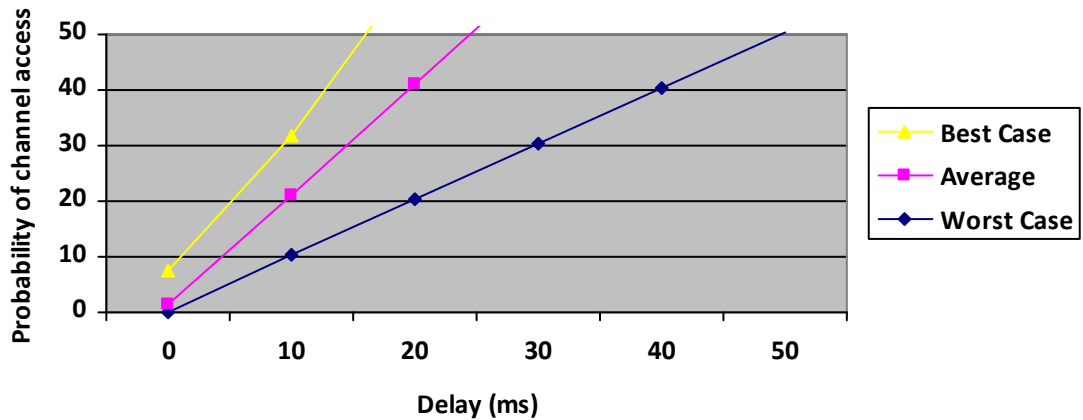

Fig 13 Channel access delay in CSMA with a sensing range of 500m, report rate 10HZ and packet length 500 byte.

When 500 bytes long packets are sent 10 times per second and the nodes have a sensing range in 1 km since this corresponds to the largest bandwidth requirements per unit area. MAC can handle 70 nodes that are in communication range of each other without packet collisions. Simulation contains situations that are overloaded and a node has 210 neighbors within communication range when the range is 1km, and consequently some packet drops takes place.

Cumulative distribution functions (CDFs) for the channel access delay is

$$F_{Tacc}(x)=Pr\{T_{acc}<x\} \quad \textbf{Eqn 6},$$

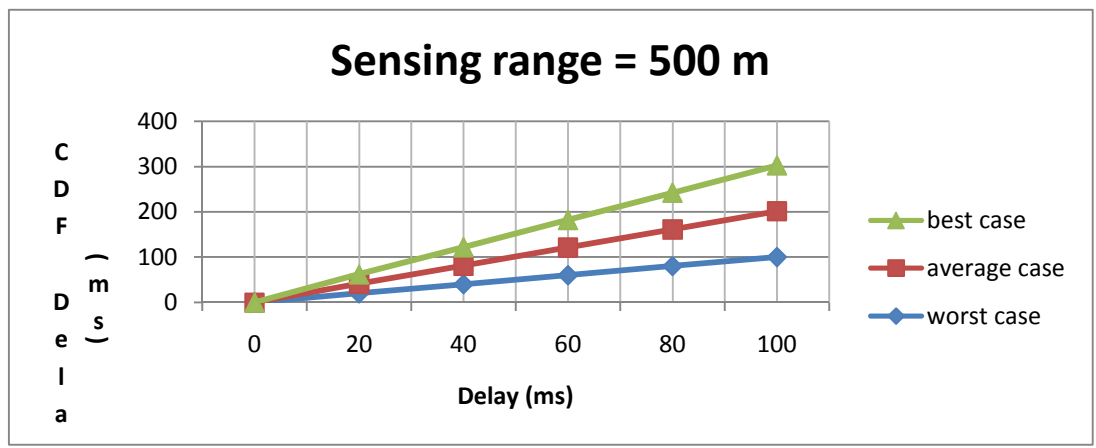

Fig. 14 - a. Sensing Range 500 meters

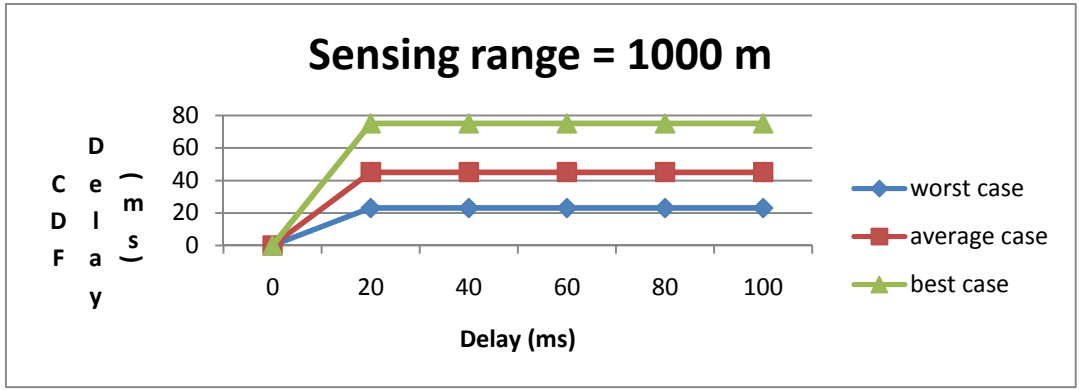

Fig. 14 b Sensing Range 1000 meters

for CSMA as in fig 14 (a,& b) for different sensing ranges. Simulation statistics were collected from middle of the of the highway with the vehicle traffic. Dropped packets are considered to have infinite delays. Three plots in the figure represent CDF for the node performance in best average and worst case for different sensing range. In best case only 5% of generated and send packets are dropped while in worst case 65% to 70% packets are dropped for sensing range of 500 meters and 50% to 55% packets are dropped in average case for sensing range of 1000 meters. Lose of many consecutive packets which will make the node invisible to the surrounding vehicles for a period of time. CDF for number of consecutive packet drops is in fig 18. For different sensing ranges. In worst case a node can drop 100 consecutive packets, implying invisibility for over 10 seconds.

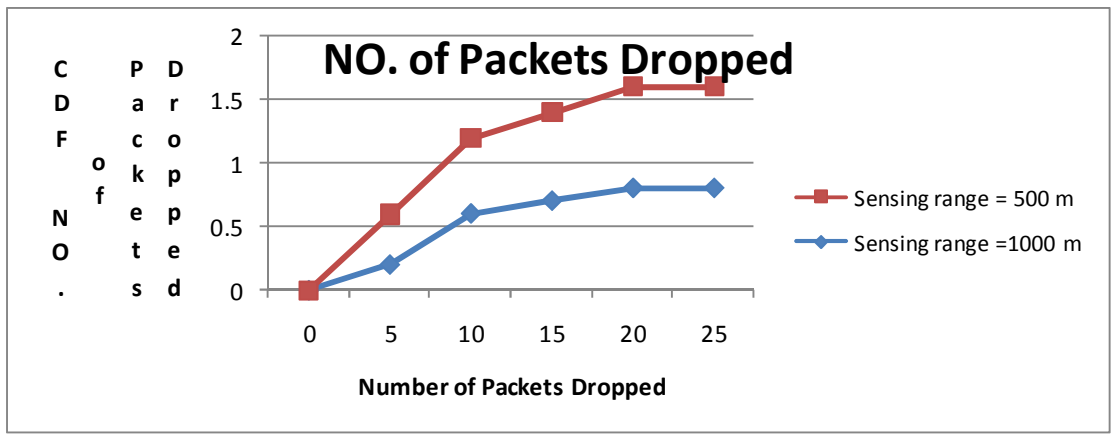

Fig. 15 Number of Packets dropped due to no channel access

The STDMA algorithm will always ensure that a vehicle node requesting channel access will be granted channel access and thus no packets are dropped. If all slots within SI are occupied, the node searching for a new node will select a slot belonging to another node. Since a vehicle node using STDMA always achieves channel access by sharing a slot with a node located far away, it is instead interesting to see how many slots that are reused in this way and how far away nodes sharing a slot are. Simulations have been carried out with the same parameter settings found in table 2. The STDMA frame size of 1 s was kept constant while the number of slots changed for different packet sizes. The results from the STDMA simulations are found in table 5, where the percentage of slots being reused within sensing range is tabulated. In the case with a sensing range of 1000 meter and 10 Hz data rate, 30% of all slots are reused within sensing range. The average distance between two vehicle nodes utilizing the same slot is approx. 800 meters. The number of nodes within sensing range is the same as in the CSMA case i.e. 230 nodes for 1000 meter and 115 nodes for 500 meter.

| STDMA | | Sensing range | | | |
|---|---|---|---|---|---|
| | | 500 meter | | 1000 meter | |
| Data Packet Rate | | 5Hz | 10 Hz | 5Hz | 10 Hz |
| Packet length | 100 byte | 0% | 0% | 0% | 0% |
| | 300 byte | 0% | 0% | 0% | 0.1% |
| | 500 byte | 0% | 1% | 0% | 30% |

Table 5.STDMA results in terms of slot reuse.

STDMA algorithm grants packets channel access since slots are reused if all slots are currently occupied within selection interval of the node. Node will choose the slot that is located furthest away hence there will be no packet drops at sending side when using STDMA and channel delay is small. Fig 15 the CDF channel delay for STDMA for all nodes will choose a slot for transmission during selection interval therefore CDF for Tacc in STDMA is sending at unity after a finite delay compared to CDF for Tacc in CSMA.

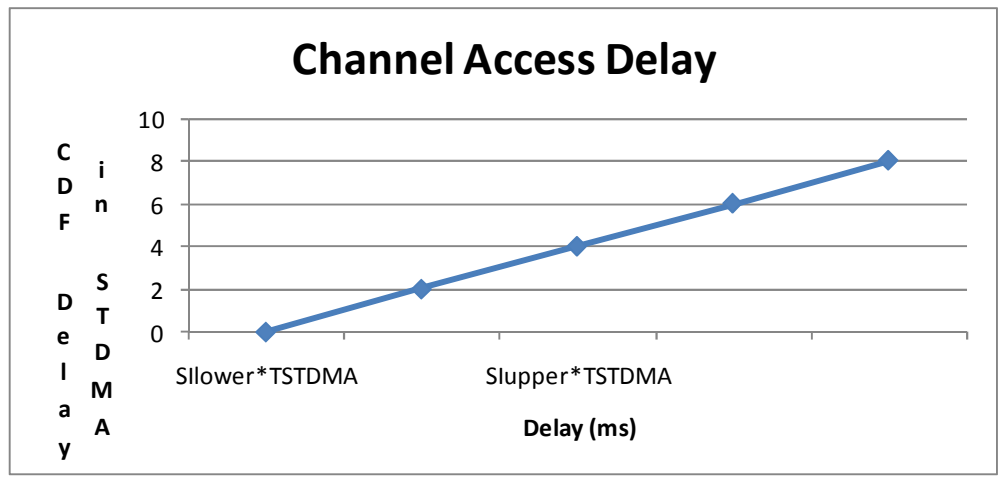

Fig. 16 CDF for channel access delay in STDMA

Fig.17 the CDF for the minimum distance between two nodes which utilizing the same slot within the sensing range is depicted for different packet lengths. When smaller packets size more nodes can be handled by the network. When long packets are used, the distance between two nodes intentionally reusing the same slot is reduced. In CSMA/CA, all channel requests did not make it to a channel access and then nodes drop packets. In CSMA/CA there is risk when nodes gets the channel access someone else also sends the packet and collision occurs. This is due to the fact that nodes can experience the channel idle at the same time, or ongoing transmission is not detected.

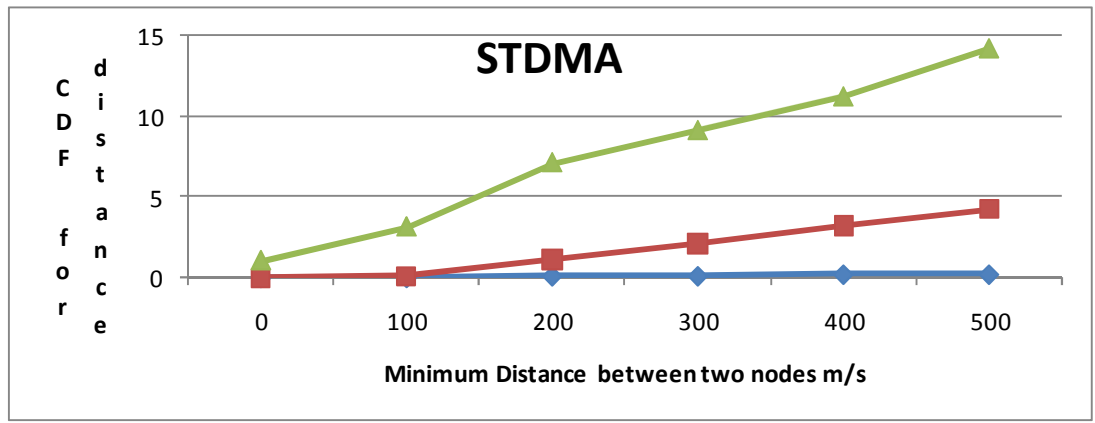

Fig. 17. Utilizing the same time slot in STDMA to find minimum distance between two nodes.

Fig. 18 the CDF for minimum distance between two nodes in CSMA/CA highway scenario sending at the same time for three different packets lengths with different ratio as shown in table above. The minimum distance can be interpreted as the distance between the nodes whose packets will, on the average, interfere the most with each other. 500 bytes, 1km sensing range scenario, about 47% of the channel requests were granted and hence we conclude that the transmitted packets will be interfered by another transmission within 500 meters in 53% of the cases.

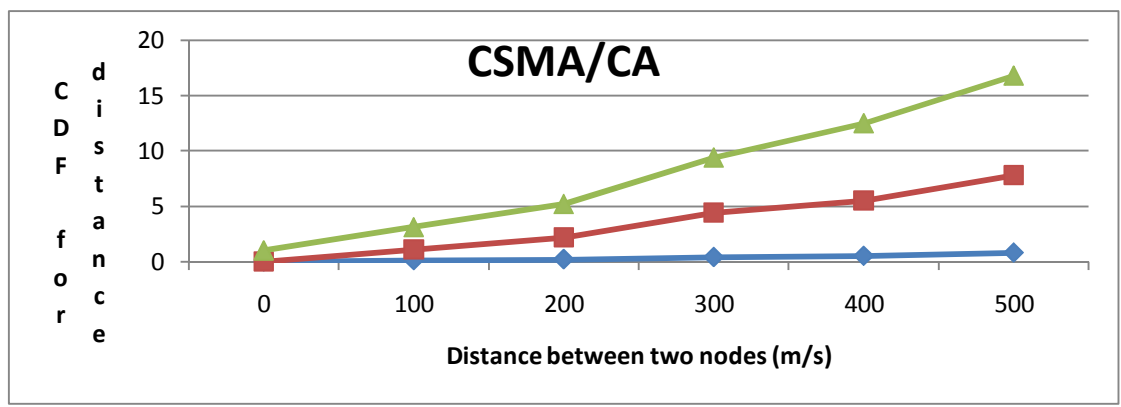

Fig. 18 Sending at the same time in CSMA/CA using 500 bytes packets.10 Hz, sensing range 1km.

# 5 CONCLUSIONS

In future traffic safety system can be classified as real-time systems which mean that the data traffic sent on the wireless channel has a deadline. The most important component of a real-time vehicle-to-vehicle communication system is the MAC protocol method. In this paper, two MAC methods have been evaluated according to their ability to meet the real-time communication deadlines. The MAC of the vehicular communication standard IEEE 802.11p CSMA was examined through simulation, and the results indicate severe performance degradation for a heavily loaded system, both for individual nodes and for the system. The simulations show that 802.11p is not suitable for periodic location messages in a Mumbai-pune highway road scenario, if the network load is high since some nodes will drop over 80% of their data packets. Location messages will be a central part of vehicle communication systems and much traffic safety application will depend on locations. The simulation results indicate how 802.11p should be configured in order to avoid severe performance loss; short packet lengths together with a low frequency range. It should be noted though that if retransmissions are used to increase reliability, the system will be heavily loaded already at low frequencies. The main drawback with CSMA is its unpredictable behavior. This implies that CSMA is unsuitable for real-time vehicle-to-vehicle communication data traffic. STDMA algorithm scheme will always have grant channel access to the number of vehicle nodes. If all slots are occupied, a vehicle node will use the same slot as another vehicle node which is situated far away from it. The worst case access time in STDMA is thus bounded and equal to the listening period plus a nominal increment. From a sending perspective STDMA outperforms CSMA during high utilization periods. The reuse of slots in STDMA is not noticeable until 500 byte long packets and an inter-arrival time of 100ms with a sensing range of 1000 meter are used. Then 30% of all slots are reused within sensing range implying a potential increase in interference, but no data packet drops. This is much better then CSMA algorithm using the same data traffic model since increased interference can be combined with coding and diversity, but the 53% packet drops in the corresponding CSMA scenario are lost.

# REFERENCES:-

## Authors

Prof. Vaishali Dinesh Khiarnar

She is a research scholar in Institute of Technology under Nirma University- Ahmadabad under theguidance of Dr. Ketan Kotecha. Her areas of interest are wireless communication, VANET, Storage etc. She has total 13 years of teaching experience and currently working as HOD of IT department in Terna Engineering College under Mumbai University 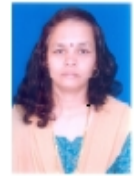

Dr. Ketan Kotecha

He is Director of Institute of Technology under Nirma University- Ahmadabad. He has total 14 years of teaching experience and he has guided several students for PG and Ph.D students. 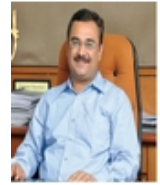